\documentclass[twocolumn]{aastex631}

\usepackage{amsmath}
\usepackage{physics}

\begin{document}
\title{How Low Can You Go: Constraining the Effects of Catalog Incompleteness on Dark Siren Cosmology}

\author[0000-0002-7456-2167]{Madison VanWyngarden}
\affiliation{Canadian Institute for Theoretical Astrophysics, University of Toronto, 60 St. George Street, Toronto, ON. M5S 3H8, Canada}
\affiliation{Steward Observatory, University of Arizona, 933 N Cherry Ave, Tucson, AZ 85712, USA}
\author[0000-0002-1980-5293]{Maya Fishbach}
\affiliation{Canadian Institute for Theoretical Astrophysics, University of Toronto, 60 St. George Street, Toronto, ON. M5S 3H8, Canada}
\affiliation{David A. Dunlap Department of Astronomy and Astrophysics, University of Toronto, 50 St. George St., Toronto ON. M5S 3H4, Canada}
\affiliation{Department of Physics, University of Toronto, 60 St. George St., Toronto, ON M5S 3H8, Canada}
\author[0000-0002-4103-0666]{Aditya Vijaykumar}
\affiliation{Canadian Institute for Theoretical Astrophysics, University of Toronto, 60 St. George Street, Toronto, ON. M5S 3H8, Canada}
\author[0000-0002-8304-0109]{Alexandra G. Guerrero}
\affiliation{Department of Physics, University of Chicago, Chicago, IL 60637, USA}
\author[0000-0002-0175-5064]{Daniel E. Holz}
\affiliation{Department of Physics, University of Chicago, Chicago, IL 60637, USA}
\affiliation{Department of Astronomy \& Astrophysics, Enrico Fermi Institute, Kavli Institute for Cosmological Physics, University of Chicago, Chicago, IL 60637, USA.}
\affiliation{NSF-Simons AI Institute for the Sky (SkAI), 172 E. Chestnut St., Chicago, IL 60611, USA}

\begin{abstract}
Gravitational waves (GWs) serve as standard sirens by directly encoding the luminosity distance to their source. When the host galaxy redshift is known, for example, through observation of an electromagnetic (EM) counterpart, GW detections can provide an independent measurement of the Hubble constant, $H_0$. However, even in the absence of an EM counterpart, inferring $H_0$ is possible through the dark siren method. In this approach, every galaxy in the GW localization volume is considered a potential host that contributes to a measurement of $H_0$, with redshift information supplied by galaxy catalogs. Using mock galaxy catalogs, we explore the effect of catalog incompleteness on dark siren measurements of $H_0$. We find that in the case of well-localized GW events, if GW hosts are found in all galaxies with  host halo masses $M_h > 2 \times10^{11} M_{\odot}h^{-1}$, catalogs only need to be complete down to the 1\% brightest magnitude $M_i < -22.43$ to draw an unbiased, informative posterior on H0. We demonstrate that this is a direct result of the clustering of fainter galaxies around brighter and more massive galaxies. For a mock galaxy catalog without clustering, or for GW localization volumes that are too large, using only the brightest galaxies results in a biased $H_0$ posterior. These results are important for informing future dark siren analyses with LIGO-Virgo-KAGRA as well as next-generation detectors.  
\end{abstract}

\section{Introduction} \label{sec:intro}
The 4-6$\sigma$ discrepancy between low-redshift measurements of  the Hubble constant, $H_0$, (e.g. supernovae and cepheids; \citealt{Scolnic2022,Riess2021,Uddin2023}) and high-redshift measurements (e.g. cosmic microwave background; \citealt{Planck2020}) has given rise to the current crisis in cosmology known as the Hubble tension \citep{Freedman2023}. Whether these differences are a result of systematics in the cosmic distance ladder or a breakdown of $\Lambda$CDM cosmology is still under investigation. Alternative cosmological probes may be able to reduce the tension or provide insight into its cause~\citep{Moresco:2022phi}. 

Decades before the first gravitational wave (GW) detection \citep{2016PhRvL.116f1102A}, \cite{Schutz1986} proposed using GW observations of compact binary mergers to measure the Hubble constant. GW signals are referred to as ``standard sirens" because their waveform directly encodes the luminosity distance to the source~\citep{HolzHughes2005}. However, their redshift is degenerate with the source-frame mass, making an independent measurement of the host galaxy's redshift essential.  When this redshift is available, a measurement of $H_0$ can be made without the need of a cosmic distance ladder \citep{Schutz1986, HolzHughes2005}. 
Bright sirens---GW sources with an electromagnetic (EM) counterpart---allow for identification of a unique host galaxy redshift. The identification of NGC 4993 as the host galaxy of the binary neutron star merger GW170817~\citep{Abbott2017a} allowed for the first measurement of $H_0$ from GWs \citep{Abbott2017b, Abbott2017d, Coulter2017, SoaresSantos2017}. However, no other GW signals with EM counterparts have been confidently confirmed. 

There are different methods for estimating the redshifts of GW events without EM counterparts. These include the spectral siren, which uses the distribution of source-frame masses to infer cosmology \citep{Taylor2012, Farr2019, PhysRevD.104.062009, Ezquiaga2022}; love sirens, which uses the neutron star equation of state \citep{Messenger2012, Chatterjee2021}; stochastic sirens, which uses the amplitude of the stochastic gravitational wave background \citep{Cousins2025}; and and approaches that utilize cross-correlations between GW sources and large-scale structure \citep{Namikawa2016,Bera2020, Mukherjee2021,2024ApJ...975..189M}. 

In this work, we focus on \cite{Schutz1986}'s original proposal, which utilized GWs without an EM counterpart, known as dark sirens, and redshift information from galaxy catalogs. In the so-called ``\textit{dark siren} method," every galaxy in the GW localization volume is considered a potential host that provides redshift information contributing to a measurement of $H_0$ \citep{DelPozzo2012, Chen:2017rfc, Fishbach2019, SoaresSantos2019, Gray2020, Palmese2020, Abbott2023, Gair2023, Palmese2023,PhysRevD.108.042002,Gray_2023}. The $H_0$ measurement is obtained by combining results from each potential host galaxy. Though each dark siren measurement is less informative than a single bright siren measurement, the number of dark sirens far exceeds the number of bright sirens, allowing for measurements to be stacked over time to increase  precision. The best localized events from observing runs O1-O4a from the LIGO-Virgo-KAGRA collaboration (LVK) has constrained $H_0$ to $\sim20\%$ uncertainty \citep{GWTC42025, Bom2024}. Dark siren cosmology will continue to play an important role with the inclusion of next-generation detectors, such as LISA \citep{Auclair2023}, Cosmic Explorer \citep{Evans2021}, and Einstein Telescope \citep{Maggiore2020, Branchesi2023}, which will greatly increase the number of dark sirens.

A persistent problem in dark siren cosmology is the incompleteness of galaxy catalogs. A GW event's localization volume can contain hundreds of galaxies for relatively well-localized events to $\mathcal{O}(10^4-10^5)$ galaxies for a typical detection \citep{Fishbach2019, Moresco2022, Chen2016}. Galaxy catalogs are inherently magnitude-limited and sky coverage and depth vary dramatically from survey to survey. For any dark siren analysis, it is possible that the true host of a GW event is not included in the galaxy catalog or catalogs overlapping with the localization volume. In this case, the host providing support for the correct value of $H_0$ would be lost. However, due to the clustering of galaxies, another in-catalog galaxy may still be present at the correct redshift \citep{MacLeod2008}. Previous studies have exploited this fact to develop completeness corrections in which the missing galaxies in an incomplete catalog are assumed to be preferentially close to the in-catalog galaxies, leveraging large-scale structure to improve dark siren analyses~\citep{2021JCAP...08..026F,dalang2024largescalestructureprior, Dalang_2024, Leyde_2024,leyde2025cosmiccartographyiicompleting,borghi2025echoesdarkgalaxycatalog}. Other studies have explored the effects of catalog incompleteness on constraining $H_0$ with real data from GWTC-3, including \cite{Naveed2025} and \cite{Beirnaert2025} which investigated the use of bright galaxies and galaxy cluster catalogs, respectively.

In this work, we further examine the effects of galaxy catalog incompleteness on the measurement of $H_0$; however, unlike previous studies, we do not apply any completeness correction and use a mock galaxy catalog to systematically investigate different limits of catalog completeness. We interpret and explain our results in the context of galaxy clustering. In Section \ref{sec:Methods}, we describe the statistical framework and mock data used in this study. In Sections \ref{sec:1LOS} and \ref{sec:200LOS}, we examine the results of using incomplete catalogs for both a simple test case and a more realistic scenario. We discuss the impact of galaxy clustering on our results in Section \ref{sec:LSS} before discussing caveats to our results in Section \ref{sec:Discussion} and presenting our conclusions in Section \ref{sec:Conclusion}. 

\section{Methods}\label{sec:Methods}
\subsection{Statistical Framework}
This section outlines the statistical framework used in our analysis, following the procedure and code provided by \cite{Gair2023}. We begin by assuming that for a given set of GW signals, we observe a set of luminosity distances $\hat{d_L}$. We can then derive a posterior on $H_0$ using Bayes' theorem,
\begin{equation}
    p(H_0|{\hat{d_L}}) \propto \mathcal{L}({\hat{d_L}}|H_0) \ p(H_0)
\end{equation}
where $p(H_0)$ is the prior on $H_0$ and  $\mathcal{L}({\hat{d_L}}|H_0)$ is the likelihood of observing $\hat{d_L}$ given a value of $H_0$. The likelihood for a single GW event is described as
\begin{equation}
    \mathcal{L}({\hat{d_L^i}}|H_0) =  \frac{\int dz \mathcal{L_{\mathrm{GW}}}(\hat{d_L^i}|d_L(z, H_0)) \ p_{\mathrm{CBC}}(z)}{\int dz P_{\mathrm{det}}^{\mathrm{GW}}(z, H_0) \ p_{\mathrm{CBC}}(z)}
\end{equation}
where $\mathcal{L_{\mathrm{GW}}}(\hat{d_L^i}|d_L(z, H_0))$ is the likelihood of measuring a GW source with observed luminosity distance $\hat{d_L^i}$ given its true luminosity distance $d_L(z, H_0)$, $p_{\mathrm{CBC}}(z)$ is the probability that the compact binary coalescence (CBC) occurred at redshift $z$ and $P_{\mathrm{det}}^{\mathrm{GW}}(z, H_0)$ is the GW detection probability. 
For our simulations, we make the simplifying assumption that the GW distance uncertainties are Gaussian, so that the GW likelihood can be written as, 
\begin{equation}\label{eq:GWlike}
    \mathcal{L}_{\mathrm{GW}}(\hat{d_L^i}|d_L(z, H_0))=\frac{1}{\sqrt{2\pi}\sigma_{d_L}}\mathrm{exp}\qty[-\frac{1}{2}\frac{(\hat{d_L^i}-d_L(z, H_0))^2}{\sigma_{d_L}^2}],
\end{equation}
where $\sigma_{d_L} = Ad_L(z, H_0)$ and $A$ is a constant fractional error on the luminosity distance. In the context of dark siren cosmology, $p_{\mathrm{CBC}}(z)$ is the redshift distribution of potential host galaxies in our catalog. We assume that all galaxy redshifts are perfectly known, so $p_{\mathrm{CBC}}(z)$ is given by a sum of Dirac delta functions, $\frac{1}{N_{\mathrm{gal}}}\sum_{i}^{N_{\mathrm{gal}}}\delta(z-z_{\mathrm{gal}}^i)$. This simplifies the single-event likelihood to the following: 
\begin{equation}
    \mathcal{L}_({\hat{d_L^i}}|H_0) = \frac{\sum_{i}^{N_{\mathrm{gal}}} \mathcal{L_{\mathrm{GW}}}(\hat{d_L^i}|d_L(z, H_0))}{\sum_{i}^{N_{\mathrm{gal}}}P_{\mathrm{det}}^{\mathrm{GW}}(z, H_0)}.
\end{equation}
Our simulations also adopt a simplified GW detection probability of 
\begin{equation}\label{eq:GWdet}
\begin{split}
    P_{\mathrm{det}}^{\mathrm{GW}}(z, H_0) & = \int_{-\infty} ^{\infty}\Theta(\hat{d_L};\hat{d_L^{\mathrm{thr}}})\mathcal{L}_{\mathrm{GW}}(\hat{d_L}|d_L(z, H_0))d\hat{d_L} \\
    & = \frac{1}{2}\qty(1+\erf\qty[\frac{d_L(z, H_0)-\hat{d_L^{\mathrm{thr}}}}{\sqrt{2}Ad_L(z, H_0)}]),
\end{split}
\end{equation}
where
$\Theta(\hat{d_L};\hat{d_L^{\mathrm{thr}}})$ is a Heaviside step function dropping to 0 for $\hat{d_L}>\hat{d_L^{\mathrm{thr}}}$. As in \cite{Gair2023}, a GW source is detected if its measured luminosity distance is positive and less than our distance threshold, $\hat{d_L^{\mathrm{thr}}}$, of $1550 \, \mathrm{Mpc}$ ($z\approx0.3$ for a standard cosmology). The distance threshold must be applied to the observed luminosity distance rather than the true luminosity distance \citep{Essick2024}. Due to fluctuations in detector noise, GW sources with true $d_L$ above the threshold value may still be detected.

\subsection{Mock Data} \label{sec:MICEcat}
This work uses MICECAT, a mock galaxy catalog produced by the MICE collaboration from the MICE grand challenge light cone halo and galaxy catalog \citep{Fosalba2015a, Crocce2015, Fosalba2015b, Carretero2015, Hoffmann2015}.  The catalog was generated using Halo Occupation Distribution and Halo Abundance Matching prescriptions to populate Friends of Friends dark matter halos with the following input cosmological parameters: $H_0= 70\, \mathrm{km/s/Mpc}$, $\Omega_m=0.25$, $\Omega_b=0.044$, $\Omega_{\Lambda}=0.75$. It is a lightcone covering one octant of the sky out to a redshift of 1.4, containing over 205 million galaxies, and is complete down to an absolute magnitude of $M_r<-18.9$ with host halo masses $M_h > 2.2 \times10^{11} M_{\odot}h^{-1}$.

To simulate dark sirens, we first generate lines of sight (LOS) within the MICECAT catalog to populate with GW events. We begin by selecting 200 random RA and Dec pairs within the MICECAT range of 0 to 90 degrees. Lines of sight are then drawn by selecting all galaxies in the catalog within a 1-degree radius from a given RA and Dec pair  ($\sim 3$ $\mathrm{deg}^2$ sky localizations). These lines of sight contain an average of $\sim 120,000$ galaxies out to $z = 1.4$. 

We note that our 1-degree opening angle is smaller than typical GW sky localizations. For CBC candidates from observing runs O1 through O4a only $\sim 10\%$ had sky localizations smaller than $100\,\mathrm{deg}^2$ \citep{GWTC42025}. To explore this regime, we also test 5-degree opening angles ($\sim 75$ $\mathrm{deg}^2$ sky localizations) and recover qualitatively similar results to those reported below using the one-degree opening angles. However, events with significantly worse localizations are not informative. Although we keep the opening angle fixed throughout our analysis, we also explore the poorly-localized regime by increasing the distance measurement uncertainty $\sigma_{d_L}$ which, analogously to larger sky areas, increases the localization volume and therefore the number of galaxies contributing to the analysis. Single-degree sky localizations are expected to be achievable with next-generation detectors \citep{Maggiore2024}. A more detailed discussion of our choice in localization volume is given in Section \ref{sec:Discussion}.

We draw GW events following the procedure of \cite{Gair2023} by first selecting galaxies with a true redshift $z<1.4$ and using the input cosmological parameters to compute the true $d_L$. An observed GW distance ($\hat{d_L}$) is then drawn from a Gaussian centered on $d_L$ with $\sigma_{d_L}=Ad_L$ where A is 0.1, 0.2, or 0.3. This process is repeated until we have 200 GW events with $\hat{d_L} < \hat{d_L^{\mathrm{thr}}}$. 

\begin{deluxetable}{c c c c c}
\tablenum{1}
\tablecaption{Luminosities corresponding to the absolute magnitude cutoffs defining the brightest galaxy fractions in units of $L_\star$
\label{tab:mgas}}
\tablehead{
\colhead{}&\colhead{1\%[$L_\star$]}&\colhead{10\%[$L_\star$]}&\colhead{20\%[$L_\star$]}&\colhead{$L_\star$}
}
\startdata
$i$-band & $2.03$ & $0.98$ & $0.69$ &  $2.97\times 10^{10}$\\
$g$-band & $3.35$ & $1.51$ & $1.08$ &  $1.34\times 10^{10}$\\
\enddata
\end{deluxetable}

\subsection{Catalog Incompleteness}
Galaxy catalogs are inherently magnitude-limited, so the faintest galaxies in any region may not be observed. 
The incompleteness of realistic galaxy catalogs is often governed by the galaxies' apparent magnitudes, and the incompleteness fraction therefore varies with redshift.
For example, the GLADE+ catalog used by the LVK collaboration for cosmological analysis from GWTC-3 and GWTC-4 is complete up to 47 Mpc, $\sim 55\%$ complete at 130 Mpc, and $\sim 20\%$ complete at 800 Mpc \citep{Dalya2022, Abbott2023, 2025arXiv250904348T}. They define completeness as the fraction of total expected B-band luminosity found in the catalog for different $d_L$ limits. Here, we are interested in simulating a fixed completeness fraction at all redshifts, so we use absolute magnitude thresholds to remove galaxies from the catalog. 

\begin{figure}[t]
    \centering
    \includegraphics[width=0.5\textwidth]{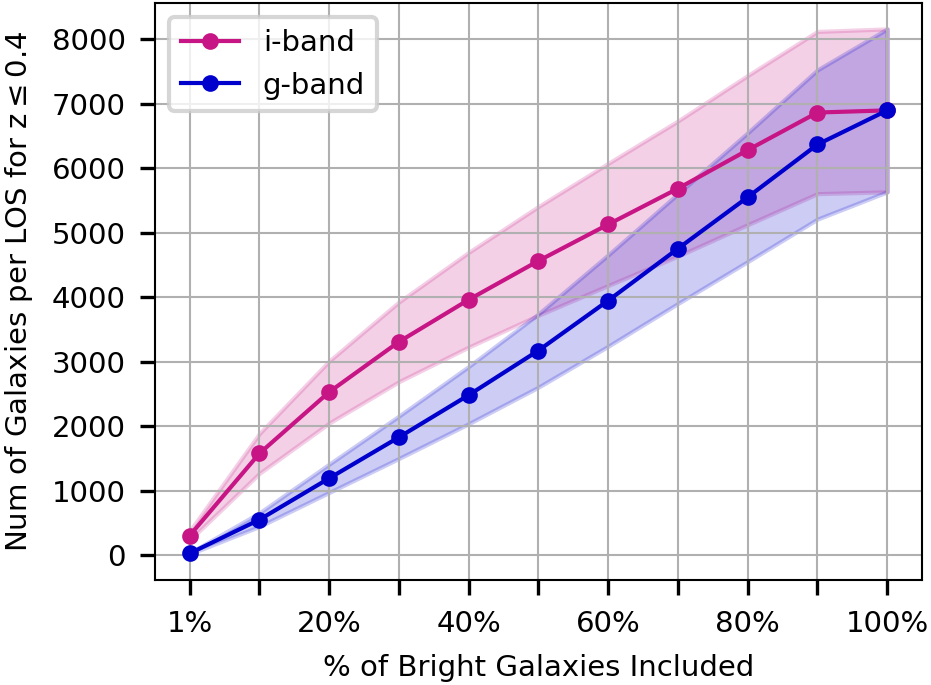}
    \caption{Average number of galaxies per $3\ \mathrm{deg}^2$ LOS for $z\leq0.4$, shown as a function of increasing brightness cutoffs in the $i$-band and $g$-band. The shaded region illustrates the standard deviation over the 200 LOS. There are greater numbers of bright $i$-band galaxies at this redshift range than bright $g$-band galaxies.}
    \label{fig:galnums}
\end{figure}

MICECAT reports galaxy magnitudes in the DES photometric bands ($grizY$). We convert these apparent magnitudes to absolute magnitudes using MICECAT's input cosmological parameters. To test the extreme case of incompleteness, along each LOS we retain only the 1\%, 10\%, and 20\% brightest galaxies (smallest absolute magnitudes) in the $i$- and $g$-band, which roughly trace stellar mass and recent star formation rate, respectively~\citep{2003ApJS..149..289B}. We find that the 1\% brightest galaxies in the $i$-band, for example, contain $ \sim 5\%$ of the total luminosity in the catalog. Retaining the 1\% brightest $i$-band galaxies is analogous to 10\% completeness under the GLADE+ definition of completeness. Using the 10\% and 20\% brightest $i$-band galaxies is analogous to $\sim 28\%$ and $\sim 44\%$ completeness under their definition. 

The luminosities corresponding to the magnitude cutoffs used to define the brightness fractions are shown in Table \ref{tab:mgas}, given in units of characteristic luminosity $L_\star$. We calculate $L_\star$ for each photometric band from the characteristic magnitudes, $M_\star$, reported in \cite{Blanton2003}. The average number of galaxies at $z\leq0.4$ per LOS for our 200 LOS is shown in Figure \ref{fig:galnums} for increasing brightness cutoffs. There are consistently more bright $i$-band galaxies at our redshifts of interest than bright $g$-band galaxies, which favor higher redshifts. This is consistent with the picture that most of the star formation in the Universe happens at $z\approx2$~\citep{2014ARA&A..52..415M}.

\begin{figure}[t]
    \centering
    \includegraphics[width=0.5\textwidth]{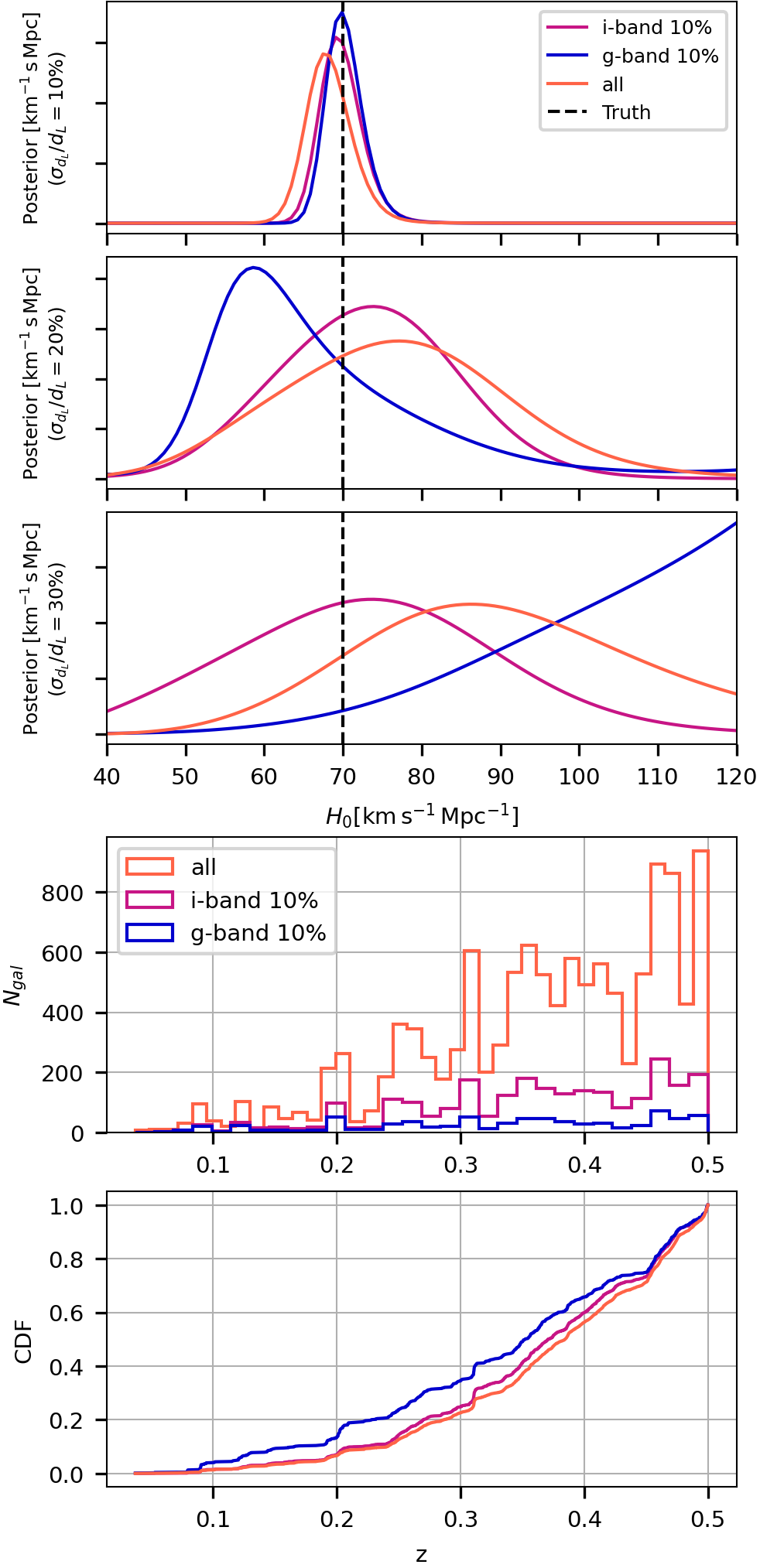}
    \caption{Posterior on $H_0$ for 200 GW events injected along one LOS and reconstructed with only the 10\% brightest galaxies in the $i$- and $g$-band with 10\%, 20\%, and 30\% error on GW luminosity distance. In the bottom panels are the normalized redshift distributions and the cumulative density function of the full LOS and the 10\% brightest $i$- and $g$-band galaxies.}
    \label{fig:1LOS}
\end{figure}

\section{Incompleteness Along a Single Line of Sight} \label{sec:1LOS}
We begin with the test case of many GW events along a single line of sight. Though this is not realistic for GW detections, it provides a simple test case to examine the influence that the specific LOS has on the $H_0$ inference. We randomly draw 200 GW host galaxies from the full line of sight, assuming each galaxy is equally likely to host a GW event. However, we select only those galaxies with magnitudes less than the 10\% cutoff shown in Table \ref{tab:mgas} to reconstruct the line of sight galaxy redshift distribution for $H_0$ inference. This means that only the 10\% brightest galaxies are used for $N_{\mathrm{gal}}$ and $d_L(z,H_0)$ in the likelihood (see Equation \ref{eq:GWlike}).  

The posteriors on $H_0$ for $\sigma_{d_L}/d_L = 0.1$, $0.2$, and $0.3$ are shown in Figure \ref{fig:1LOS}. Increasing the error on $d_L$ is analogous to increasing the size of the GW localization volume. As expected, the posteriors widen and become less informative as the localization volume increases. 
The posteriors constructed with only the 10\% brightest galaxies are remarkably similar to the posterior constructed with the full galaxy sample, especially in the best localized case. Using only a subset of bright galaxies also produces posteriors that are sharper than the full posterior. The $i$-band bright posterior better mimics the full posterior in the $\sigma_{d_L}/d_L = 0.2$ and $0.3$ cases than the $g$-band bright posterior, indicating that the bright $i$-band galaxies are a better tracer of the full galaxy distribution at the redshift range of GW observations. Additionally, there are simply more bright $i$-band galaxies at these redshifts, as shown in Figure \ref{fig:galnums}. This is further confirmed by the CDF of the galaxy redshift distributions for each galaxy population shown in the bottom panel of Figure \ref{fig:1LOS}, where the 10\% brightest $i$-band galaxies are aligned with the full redshift distribution. 

As discussed in \citet{Hanselman2024}, for poorly localized GW events (large errors on $d_L$), the $H_0$ posterior becomes more sensitive to global trends in the redshift distribution (see also \citealt{2024arXiv240507904P}), which differ between the bright $i$-band and bright $g$-band galaxies.
For most LOS, bright $g$-band galaxies are preferentially found at higher redshifts, following the star-formation rate which peaks at $z\sim2$, and will therefore bias the $H_0$ posterior to higher values. Given small number statistics with this specific LOS, however, the distribution of the 10\% brightest $g$-band galaxies skews to lower redshifts.
The shapes of the posteriors are also sensitive to the galaxy clustering in this particular LOS, as over- or under-densities in the redshift distribution can cause the $H_0$ posteriors to fluctuate for individual events.
The $H_0$ inference is unbiased only if the redshift distribution of the GW events matches the galaxy redshift distribution used in the reconstruction.

Overall, we find that using only the 10\% brightest galaxies is sufficient for the dark siren measurement for well-localized GW events. In other words, there is little information lost from catalog incompleteness. Although we only show a single LOS, we repeat the inference with several LOS to verify that the qualitative trends hold. Averaging over three LOS, for the best localized case (10\% distance error), the average difference between the 10\% brightest $i$-band posterior and the `all' posterior is 0.04$\sigma$, while the average difference between the 10\% brightest $g$-band and `all' is 0.28$\sigma$. Here, the difference is defined as the difference in mean between the bright and full posterior divided by the width of the full posterior. We have chosen to express the differences in terms of $\sigma$ for clarity, but we note that the posteriors are not Gaussian. If we increase the catalog completeness and use the 20\% brightest galaxies, the average differences in the full and bright posteriors become 0.01$\sigma$ for bright $i$-band galaxies and 0.18$\sigma$ for bright $g$-band galaxies in the best-localized case. 

\begin{figure}[t]
    \centering
    \includegraphics[width=0.5\textwidth]{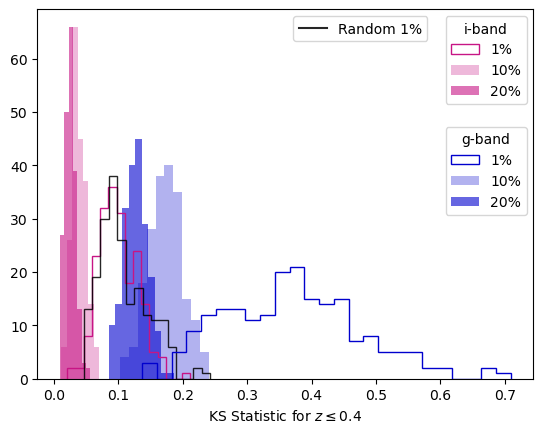}
    \caption{KS statistics between the redshift distributions of the 1\%, 10\%, and 20\% brightest galaxies and the full galaxy sample along our 200 LOS for z$\leq$0.4. We also include the KS statistics between the redshift distribution of a random 1\% of galaxies with the full LOS for comparison.}
    \label{fig:KS}
\end{figure}

The similarities in posteriors between the bright sample of galaxies and the full sample, and the difference between $i$-band and $g$-band results can be explained by how the bright galaxies trace the redshift distribution of a given LOS. Galaxies tend to cluster around the most massive, and thus brightest, galaxies. This results in bright galaxies being a good tracer of the full galaxy distribution. To demonstrate this fact, we calculated the KS statistic between the redshift distribution of the brightest galaxies compared to the full distribution along our 200 LOS for $z\leq0.4$. A histogram of the KS statistics is shown in Figure \ref{fig:KS}. 
The $i$-band galaxies at all brightness intervals have smaller KS statistics, and thus better match the full galaxy distribution than the bright $g$-band galaxies. Again, there are also more bright $i$-band galaxies at the redshifts of GW observations (see Figure \ref{fig:galnums}). From Figure \ref{fig:KS}, we see the 1\% brightest $i$-band galaxies trace the full line of sight distribution as well as a random selection of 1\% of the galaxies along the LOS (the random sample is by definition drawn from the full galaxy distribution, and would thus have the smallest possible KS statistic for the given sample size). The $g$-band galaxies are worse tracers, with the 1\% brightest $g$-band galaxies differing from the full LOS by 20-60\%. Increasing the number of $g$-band galaxies considered greatly reduces the KS statistic and LOS variation. 
\begin{figure*}[t]
\includegraphics[width=\textwidth]{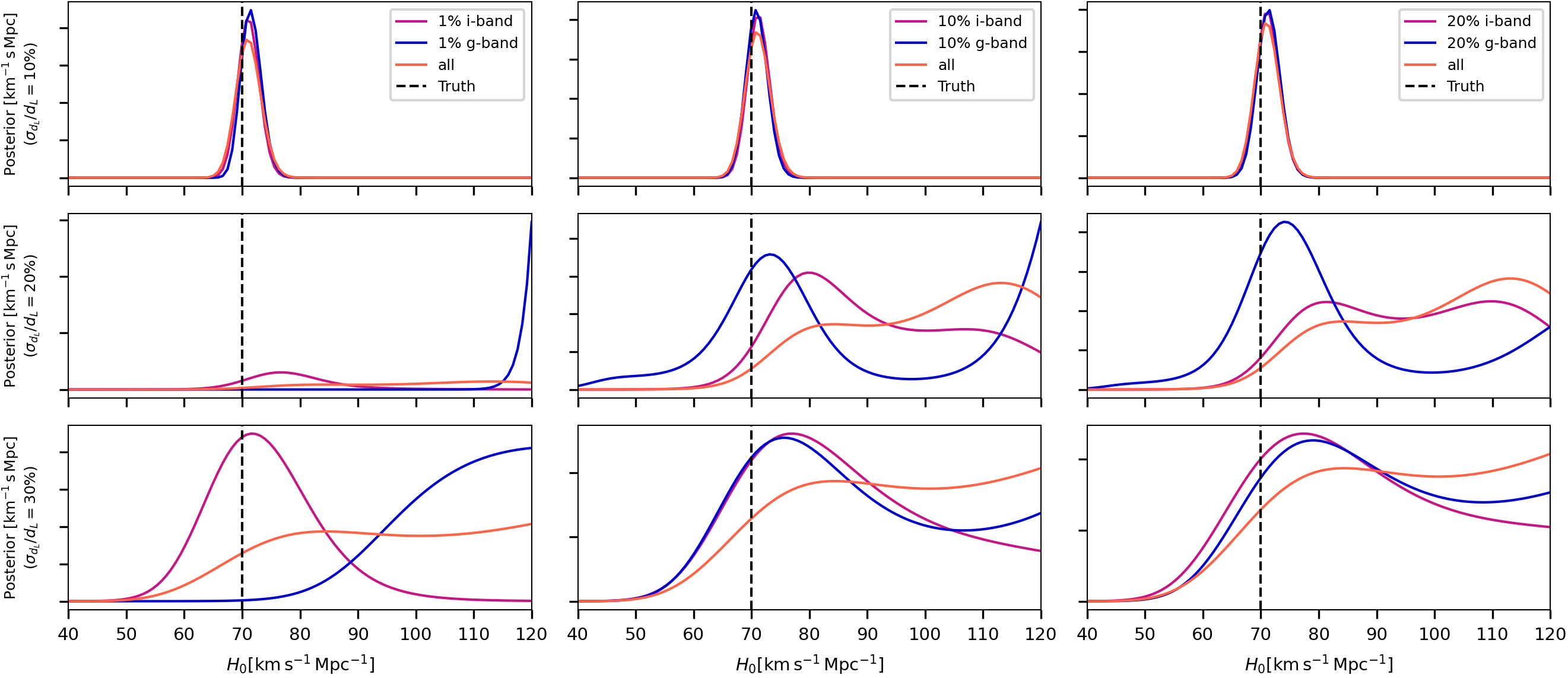}
\caption{Resulting posteriors on $H_0$ from injecting 1 GW event into each of 200 LOS and reconstructing the LOS redshift distribution with the 1\%, 10\%, and 20\% brightest galaxies in the $i-$ and $g$-band.} 
\label{fig:200LOS}
\end{figure*}

\section{Incompleteness Along Many Lines of Sight}\label{sec:200LOS}
In a more realistic analysis, GW events will be isotropically distributed instead of being concentrated along single LOS. To simulate this case, we injected one GW source into each of the 200 LOS with luminosity distance uncertainties of 10, 20, and 30\%. We reconstructed each LOS using only the galaxies with absolute magnitudes falling at or below the limits set in Table \ref{tab:mgas} for the 1, 10, and 20\% brightest galaxies, respectively. The posterior generated for each GW event is then combined to form a final posterior as shown in Figure \ref{fig:200LOS}.

We begin our analysis of these results with the best localized case at $\sigma_{d_L}/d_L = 0.1$ (top row of Figure~\ref{fig:200LOS}). The posteriors constructed from only the brightest galaxies are in strong agreement with the posteriors using all galaxies, even when retaining only the 1\% brightest galaxies in the catalog. In this regime we also see little change in the posteriors as we increase the galaxy sample from the 1\% brightest to the 20\% brightest. With small errors on the luminosity distances of our GW events, the majority of the catalog is not necessary for building an informative posterior on $H_0$. 

In the case where $\sigma_{d_L}/d_L = 0.2$, larger differences in the posteriors are observed, with the bright $i$-band posterior most closely tracing the full catalog posterior. Using only the 1\% brightest galaxies results in a $g$-band posterior that peaks sharply at the edge of the prior range. This is largely due to the fact that there are fewer $g$-band galaxies within this magnitude limit at our redshifts of interest from which to reconstruct the full galaxy distribution. The bright $g$-band galaxies that are included in the 1\% brightest sample are biased towards higher redshifts, resulting in a posterior that peaks at larger values of $H_0$. As the brightness fraction increases, the $i$-band posterior begins to converge with the full sample. The bright $g$-band posterior, on the other hand, has more support at the injected value of $H_0$ but differs significantly when compared to results obtained using all the galaxies. 

Finally, in the $\sigma_{d_L}/d_l = 0.3$ case, all posteriors have strong support at high values of $H_0$, similar to the results obtained in the single line of sight test. All three posteriors peak around the same value but the $i$ and $g$-band bright posteriors most closely resemble each other rather than the full catalog line of sight, though all three begin to converge as the brightness fraction increases. As the error on $d_L$ increases, structures in the redshift distribution begin to be washed out, so the posterior is informed by the overall redshift distribution rather than the specific large-scale structure of any LOS. These results are in good agreement with \cite{Hanselman2024}.  The posteriors in the worst localized $\sigma_{d_L}/d_L = 0.3$ regime are often less biased than the posteriors in the intermediate $\sigma_{d_L}/ d_L = 0.2$ case. This is because at the intermediate localization, there is significant support from the non-host galaxies that bias the measurement. Although these galaxies also contribute in the worst localized case, their contribution is washed out as the overall posterior is broader and therefore less biased (while being less informative).

\begin{figure}[t]
    \centering
    \includegraphics[width=0.5\textwidth]{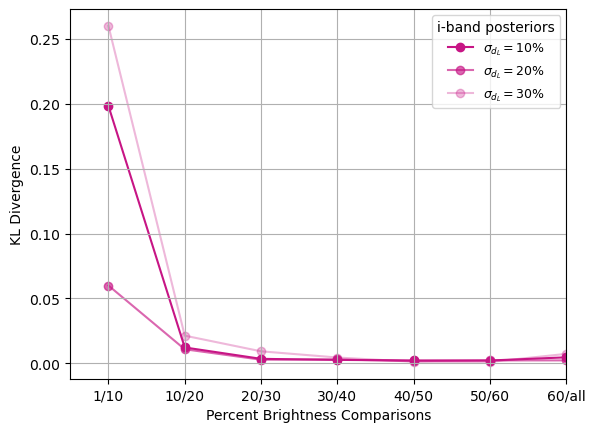}
    \caption{Median KL divergence between posteriors of increasing brightness fractions over 10 realizations of GW events. Here we show only results from posteriors constructed with the bright $i$-band galaxies at our three levels of $d_L$ uncertainty.}
    \label{fig:KL}
\end{figure}
\begin{figure*}[t]
    \centering
    \includegraphics[width=\textwidth]{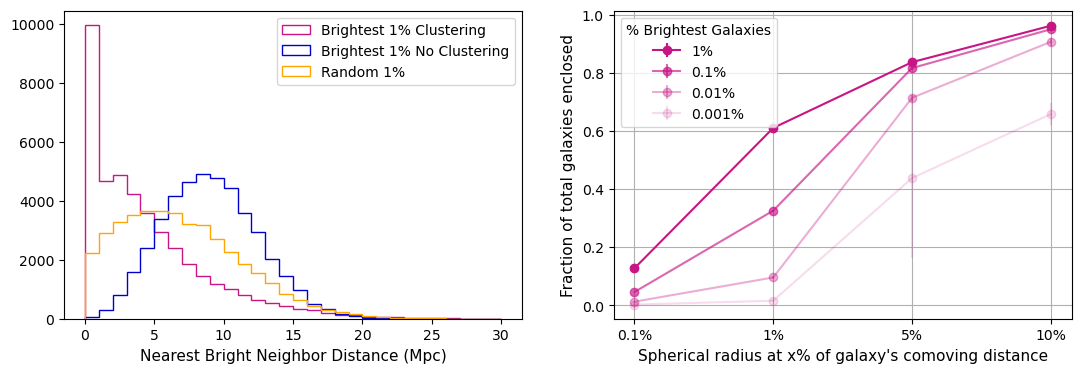}
    \caption{[\textit{Left}] Distances in Mpc to the nearest bright neighbor galaxy for the 1\% brightest $i$-band galaxies from MICECAT, the 1\% brightest $i$-band galaxies without clustering, and 1\% of galaxies drawn randomly from MICECAT, all for $z\leq0.45$. [\textit{Right}] For the 1\%, 0.1\%, 0.01\%, and 0.001\% brightest $i$-band galaxies, we draw a sphere with radius 0.1\%, 1\%, 5\%, and 10\% of that galaxy's comoving distance and calculate the fraction of total galaxies with $z\leq0.45$ enclosed by all the spheres. Results are averaged over three independent 20 square degree boxes in MICECAT.}
    \label{fig:cluster}
\end{figure*}
The influence of galaxy clustering is also diminished as the localization volume increases by increasing the sky localization. Widening our LOSs from a 1-degree opening radius to a 5-degree radius increases the widths of the $H_0$ posteriors. Using the same injected GW events and comparing at the 95\% credible interval on $H_0$ for the smallest $\sigma_{d_L}$, the posterior constructed from the 10\% brightest galaxies is a factor of 1.8 to 2.7 wider in the 5 degree LOS case for $i$-band and $g$-band, respectively. At the largest $\sigma_{d_L}$, the 10\% brightest $i$-band and $g$-band posteriors are wider by a factor of 1.3 and 1.8.

\subsection{Effects of additional galaxies}

The results of Figure \ref{fig:200LOS} prompt the question: what fraction of galaxies do we actually need in order to avoid biasing our results? Figure \ref{fig:200LOS} illustrates that as the brightness fraction increases, the bright galaxy posteriors begin to converge with the results from using all galaxies. To further investigate this effect, we calculate the Kullback-Leibler divergence (KL divergence) between the posteriors of increasing brightness fractions (e.g., the KL divergence between the $H_0$ posterior recovered with the 1\% brightest versus the 10\% brightest $i$-band galaxies) to quantify the information gained by including additional bright galaxies. This statistic measures the difference, or relative entropy, between two probability distributions. The median KL divergences for the $H_0$ posteriors constructed from the bright $i$-band galaxies are shown in Figure \ref{fig:KL} over ten realizations of the dark siren analysis described in Section \ref{sec:200LOS}. 

From Figure \ref{fig:KL} it is apparent that little information is gained after the inclusion of the 40\% brightest $i$-band galaxies. Even adding galaxies beyond the 10\% threshold leads to only a small gain in information. The limited effect of additional $i$-band galaxies can be explained by two factors. First, the bright $i$-band galaxies, even in small numbers, are good tracers of overall structure as seen in Figure \ref{fig:KS}. Second, the growth in the average number of bright $i$-band galaxies per LOS, shown in Figure \ref{fig:galnums}, decreases as larger percentages of bright galaxies are considered, because most of the brightest $i$-band galaxies in the Universe are found at low redshifts.

Though not shown in Figure \ref{fig:KL}, the posteriors constructed from the bright $g$-band galaxies exhibit a similar behavior. For the well-localized events, there is little information gain after including the 40\% brightest galaxies. However, the KL divergences between the 1\% and 10\% posteriors in the $g$-band case are much larger than the $i$-band galaxies. This likely occurs because there are far fewer $g$-band galaxies within the 1\% brightness threshold at our redshifts of interest than $i$-band galaxies. As a result,  there are realizations in which the posterior from the 1\% brightest $g$-band galaxies is entirely uninformative. For $\sigma_{d_L} > 10\%$,  there are realizations in which the $g$-band posterior fails to converge with the full posterior even after including the 60\% brightest galaxies. Again, even at 60\% brightness, there remains an average 1000 galaxy difference per LOS between the number of bright $i$-band and $g$-band galaxies. Except for the well-localized regime, completeness levels greater than 60\% are required for the $H_0$ posterior constructed with the bright $g$-band galaxies to converge to the posterior recovered from all galaxies. 

\section{Impact of Galaxy Clustering} \label{sec:LSS}
The ability of an incomplete galaxy catalog to closely reproduce the posteriors constructed from a complete catalog is due to the role of galaxy clustering.
Fainter galaxies tend to cluster around the most massive, brightest galaxies. In the case where the true GW host is missing from the catalog, the next closest bright galaxy is not far, as shown in the left panel of Figure~\ref{fig:cluster}, resulting in only a small redshift difference. For example, if the 90\% faintest galaxies at $z\leq0.4$ in the $i$-band are missing from our catalog, the fractional redshift difference between a faint galaxy and the closest (in comoving distance) bright galaxy is $\Delta z/z \leq 0.014$ for 99\% of the faint galaxies.  The 99th percentile of the fractional redshift difference between the 90\% faintest $g$-band galaxies and bright galaxies is $\Delta z/z \leq 0.011$. These fractional redshift differences introduce such little error on $H_0$ that in the well-localized regime, an incomplete catalog produces a posterior just as informative as a complete one. 

To better understand the role that galaxy clustering plays in the dark siren analysis, we construct a catalog that matches the properties of the MICECAT catalog but lacks galaxy clustering. We preserve the relations between redshift and i- and $g$-band magnitudes as in the MICECAT catalog, but randomly draw and re-assign RA and Dec of each galaxy. This scrambles the sky positions, erasing the original large-scale structure while preserving the same trends between galaxy brightness and redshift.

The effects of clustering can be clearly seen in Figure \ref{fig:cluster}. On the left, we show the nearest neighbor distances for the 1\% brightest galaxies in the $i$-band, the same population but in our catalog without clustering, and a 1\% of galaxies drawn randomly from MICECAT. The 1\% brightest MICECAT galaxies are significantly more clustered than the other two galaxy populations, evident from the peak at nearest bright neighbor distances of a few Mpc.
The plot on the right seeks to further probe just how clustered the brightest galaxies are. For a 20 square degree box in MICECAT, we begin by identifying the 1\% brightest $i$-band galaxies at $z\leq0.4$ and drawing a sphere around each bright galaxy with a radius corresponding to 10\% of that galaxy's comoving distance. We then calculate what fraction of all the galaxies in our catalog box fall within the bright-galaxy-centered spheres, considering galaxies up to $z\leq0.45$. We begin with a radius of 10\% of the galaxy's distance, motivated by the smallest distance uncertainty considered in this study. We then considered smaller and smaller circles, at 5\%, 1\%, and 0.1\% of the galaxy's distance and find that large fractions of galaxies are still captured. We also consider even smaller subsets of bright galaxies, taking the 0.1\%, 0.01\%, and 0.001\% brightest $i$-band galaxies, and find that at the largest radii tested (corresponding to typical GW localization errors), these small fractions of galaxies still capture all, or nearly all, of the galaxies in the catalog box. Figure \ref{fig:cluster} indicates that even if only a small fraction of bright galaxies are captured by a catalog, the missing galaxies are likely found within a volume corresponding to a typical GW localization volume. 

\begin{figure}[t]
    \centering
    \includegraphics[width=0.5\textwidth]{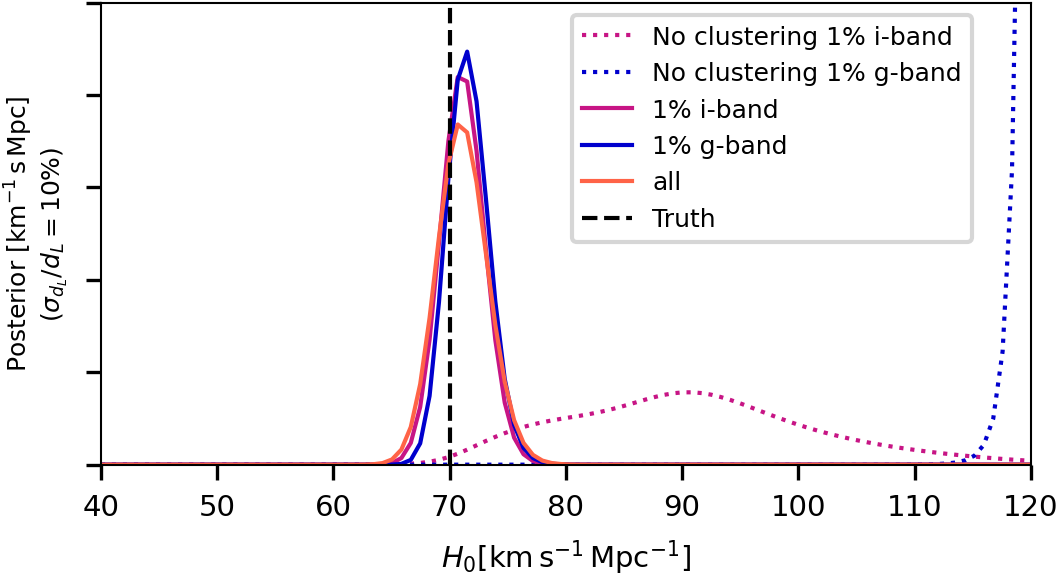}
    \caption{$H_0$ posterior constructed from the 1\% brightest galaxies in the catalog without clustering as compared to the posteriors shown in Figure \ref{fig:200LOS} for the best localized case.}
    \label{fig:uniform}
\end{figure}
The clustering evident in Figure \ref{fig:cluster} is directly responsible for the negligible effects of catalog incompleteness seen in Figure \ref{fig:200LOS} for well localized events. We further confirmed this by using our ``no-clustering" catalog for a dark siren analysis. Following the same procedure as described in Section \ref{sec:200LOS}, we inject one GW event each into our 200 LOS and construct a posterior on $H_0$ using only the 1\% brightest galaxies in the $i$ and $g$-band to reconstruct the LOS redshift distribution. The results for the best-localized case are shown in Figure \ref{fig:uniform}. Without the aid of galaxy clustering (dashed posteriors), catalog incompleteness becomes detrimental to the $H_0$ measurement. If the host is not present in the catalog, there are no longer other galaxies nearby providing support at the correct redshift. This leads to biased posteriors, and warrants methods to correct for incompleteness.

\section{Discussion} \label{sec:Discussion}
We have examined the effects of catalog incompleteness on constraining $H_0$ with the dark siren method. Our analysis relies on a number of assumptions to generate mock data. In this section, we will discuss the assumptions made and how they may impact the results presented in Sections \ref{sec:1LOS}
and \ref{sec:200LOS}.

The first assumption is that every galaxy is equally likely to host a GW source. When generating our mock data, we inject GW events randomly along each LOS. Though this choice allowed us to isolate the effects of catalog incompleteness, there is evidence that a galaxy's stellar mass or star-formation rate is positively correlated with its probability of hosting a GW event~\citep{Vijaykumar2024, Adhikari2020}. While our results indicate that a stellar-mass weighted galaxy sample (i.e. bright $i$-band galaxies) better traces the line of sight redshift distribution and is thus more useful for a dark siren analysis, these results could change if GW sources preferentially came from star-forming (bright $g$-band) galaxies. 

We also assume that redshifts are perfectly known. This is not the case in previous dark siren analyses, as photometric catalogs are often used due to their greater degree of completeness and sky coverage \citep[e.g.,][]{Abbott2023,2025arXiv250904348T}.  In this study, assuming photometric redshift errors would widen the $H_0$ posterior and introduce greater uncertainty to the measurement of $H_0$. However, photometric catalogs will not qualitatively change our results about the effects of incompleteness. In fact, the limited impact of catalog incompleteness demonstrated in this work suggests that restricting galaxy catalogs to spectroscopic surveys for future dark siren analyses may be useful to reduce the possibility of an $H_0$ bias from photometric redshift errors \citep{Turski2023}.

Additionally, our choice of 1 deg radius LOSs is optimistic for current LVK capabilities. As discussed in Section \ref{sec:MICEcat}, $\sim10\%$ of candidates from O1-O4a had sky localizations less than 100 $\mathrm{deg}^2$.  However, after the inclusion of LIGO-India in O5, the five-detector network (3 LIGOs plus Virgo and KAGRA) is expected to detect 38 BBHs with sky localizations $\leq 1 \, \mathrm{deg}^2$ at 90\% credibility and $\sim 1000$ BBHs will have $d_L$ errors of less than 10\% per year \citep{Maggiore2024}. While the best-localized regime in this analysis may not be achievable with the most recent LVK observing run, it will be with the next observing run and even more so after the construction of next-generation detectors, such as Cosmic Explorer and Einstein Telescope. 

We have demonstrated that only small fractions of the brightest galaxies in a GW localization volume are necessary for an unbiased dark siren analysis. How does this compare to the magnitude limits of real galaxy surveys? As an example, the DESI Luminous Red Galaxy Survey~\citep{2023AJ....165...58Z} has an apparent magnitude limit of  $z_{\mathrm{fiber}}<21.6$, where $z_{\mathrm{fiber}}$ is the z-band magnitude in a 1.5 arcsecond diameter centered on the galaxy. Using solar absolute magnitudes from \cite{Willmer2018}, at our threshold $d_L$, this corresponds to a luminosity cutoff of $3.47 \times 10^9 \space L_{\odot}$. This limit is far less conservative than the luminosity limits imposed in this study. Even surveys capturing only the brightest galaxies in the field could still be useful for a dark siren analysis. For example, the DESI bright galaxy survey's bright sample institutes an apparent magnitude cutoff of $r<19.5$ \citep{Hahn2023}, corresponding to a luminosity cutoff of $2.76 \times 10^{10} \space L_{\odot}$ at our threshold $d_L$. This is similar to the luminosity cutoff of our 10\% brightest sample, which provided more than enough galaxies to draw an informative posterior.

\section{Conclusion}\label{sec:Conclusion}
GWs are unique and useful cosmological probes. With the dark siren method it is possible to infer $H_0$ by treating every galaxy in the GW localization volume as a potential host. We have examined the effects of catalog incompleteness on our ability to measure $H_0$ using the dark siren method by injecting GW sources into a mock catalog, artificially removing galaxies, and completing a dark siren analysis using only the brightest galaxies along a given LOS. 

We found that for well-localized events, there is little information loss due to catalog incompleteness. If GW hosts are found in all galaxies with host halo masses $M_h > 2 \times10^{11} M_{\odot}h^{-1}$, using a catalog only complete down to the 1\% brightest absolute magnitude $M_i < -22.43$ is sufficient to draw an unbiased, informative posterior on $H_0$ in this regime. Even smaller subsets of bright galaxies are sufficient tracers of a line of sight's redshift distribution. As measured by the KL divergence, including more than the $\sim20\%$ brightest galaxies does not increase the information content of the $H_0$ posterior for well-localized events. These results are sensitive to our choice of bright $i$-band galaxies vs bright $g$-band galaxies. We found that the bright $i$-band galaxies, corresponding to greater stellar masses, were a better tracer of the full galaxy redshift distribution at the redshifts of GW observations, conservatively assuming that all galaxies are equally likely to host GW events. If GW events are preferentially found in stellar mass weighted galaxies, our conclusions would be amplified.

We showed that the efficacy of bright galaxies in a dark siren analysis is largely due to the influence of galaxy clustering. Faint galaxies cluster around the most massive and thus brightest galaxies, so if a faint GW host galaxy is missing from the catalog, there is likely a bright galaxy nearby still providing support at the correct redshift. After artificially removing the galaxy clustering from our catalog, we found that using only the brightest galaxies was no longer sufficient for drawing an informative posterior. 

We also conclude that it is only worth using the best-localized events in future dark siren analyses. As the localization volume increases, the effects of catalog incompleteness become noticeable as the $H_0$ posterior becomes more sensitive to trends in the global redshift distribution. Additionally, larger localization volumes are unable to constrain $H_0$ to precisions useful for cosmology. With the inclusion of LIGO-India to the current network of GW detectors in O5 and the potential of next-generation detectors, the number of well-localized events will greatly increase, providing a rich dataset for dark siren cosmology.

\section{Acknowledgments}
This work was completed with support from the Fulbright U.S. Student Program, sponsored by the U.S. Department of State and administered by the Institute of International Education and the Foundation for Educational Exchange between Canada and the United States of America. 
MF acknowledges support from the Natural Sciences and Engineering Research Council of Canada (NSERC) under grant RGPIN-2023-05511, the University of Toronto Connaught Fund, and the Alfred P. Sloan Foundation. AV acknowledges support from the NSERC (funding reference number 568580). DEH and AGG were supported by NSF grants PHY-2110507 and PHY-2513312. DEH was also supported by the NSF-Simons AI-Institute for the Sky (SkAI) via grants NSF AST-2421845 and Simons Foundation MPS-AI-00010513,  the Simons Collaboration on Black Holes and Strong Gravity, and the Kavli Institute for Cosmological Physics through an endowment from the Kavli Foundation.

\bibliography{mybib.bib}
\end{document}